\documentclass[sigconf]{acmart}
\pdfoutput=1
\usepackage{booktabs}
\usepackage{bm}
\usepackage{algorithm}
\usepackage{algorithmic}
\usepackage{subfigure}
\usepackage{graphicx}
\hypersetup{draft}
\copyrightyear{2019}
\acmYear{2019}
\setcopyright{acmcopyright}
\acmConference[DAC '19]{The 56th Annual Design Automation Conference 2019}{June 2--6, 2019}{Las Vegas, NV, USA}
\acmBooktitle{The 56th Annual Design Automation Conference 2019 (DAC '19), June 2--6, 2019, Las Vegas, NV, USA}
\acmPrice{15.00}
\acmDOI{10.1145/3316781.3317765}
\acmISBN{978-1-4503-6725-7/19/06}

\begin{document}

\title{An Efficient Multi-fidelity Bayesian Optimization Approach for Analog Circuit Synthesis}

\author{\vspace{-3.5mm}Shuhan Zhang$^1$, Wenlong Lyu$^1$, Fan Yang$^{1*}$, Changhao Yan$^1$, \\}
\author{\vspace{-3.5mm}Dian Zhou$^{1,3}$, Xuan Zeng$^{1*}$ and Xiangdong Hu$^{2*}$ }
\affiliation{\vspace{-3.5mm}$^1$State Key Lab of ASIC \& System, Microelectronics Department, Fudan University, China\vspace{-3.5mm}}
\affiliation{$^2$Shanghai High Performance Integrated Circuit Design Center, China \vspace{-3.5mm}}
\affiliation{$^3$University of Texas at Dallas, Dallas, USA}

\thanks{*Corresponding authors: huxd@icdc.org.cn, \{yangfan, xzeng\}@fudan.edu.cn}

\begin{abstract}

This paper presents an efficient multi-fidelity Bayesian optimization approach for analog circuit synthesis. The proposed method can significantly reduce the overall computational cost by fusing the simple but potentially inaccurate low-fidelity model and a few accurate but expensive high-fidelity data. Gaussian Process (GP) models are employed to model the low- and high-fidelity black-box functions separately. The nonlinear map between the low-fidelity model and high-fidelity model is also modelled as a Gaussian process. A fusing GP model which combines the low- and high-fidelity models can thus be built. An acquisition function based on the fusing GP model is used to balance the exploitation and exploration. The fusing GP model is evolved gradually as new data points are selected sequentially by maximizing the acquisition function. Experimental results show that our proposed method reduces up to 65.5\% of the simulation time compared with the state-of-the-art single-fidelity Bayesian optimization method, while exhibiting more stable performance and a more promising practical prospect.


\end{abstract}

\keywords{Multi-fidelity Bayesian optimization, Gaussian process, Analog circuit synthesis}

\maketitle

\section{Introduction}







As device size shrinks to the nanoscale region, the analog circuit design is facing enormous challenges due to the more and more complicated device models and expensive transistor-level simulations. The growing demands for high performance, low power and short time-to-market make manual analog circuit design even more difficult. Sophisticated automated analog circuit design is in great demand and has attracted lots of attention in both industry and academic community \cite{rutenbar2007hierarchical}.

Analog circuit simulations are computation intensive. Great efforts have been made to reduce the simulation time while searching for the global optimum efficiently. Generally, the optimization algorithms can be classified into two categories: the model-based and simulation-based approaches. Driven by simplified equations or polynomials, the model-based methods build models to approximate the performance of the analog circuit, so as to fully search the solution space without expensive computational cost. One modelling technique that has been studied in depth is the geometric programming based approach \cite{boyd2007tutorial, del2002design, colleran2003optimization}. However, a fundamental limitation of the model-based approach is that the accuracy of the constructed model is not guaranteed, and the model building process itself also requires lots of simulations. Simulation-based approaches simply take the objective function as a black-box function and activate the simulation process online. In order to better explore the solution space, a variety of well-developed global optimization algorithms have been consecutively proposed, examples include evolutionary algorithm \cite{liu2009analog}, particle swarm optimization algorithm \cite{vural2012analog}, multiple starting point optimization algorithm \cite{yang2018smart, peng2016efficient}, and simulated annealing algorithm \cite{phelps2000anaconda}. The main disadvantages that prevent the simulation-based approach from widespread use are its relatively low convergence rate and the corresponding large simulation costs.

To chart a way forward out of the dilemma, the GP-based Bayesian optimization (BO) approach has been proposed recently to combine both the model-based and simulation-based approach \cite{lyu2018efficient}. Generally, Bayesian optimization approach consists of two key elements: the surrogate model and the acquisition function \cite{shahriari2016taking}. The surrogate model is trained to mimic the behavior of the latent function. And the acquisition function helps to select the next query point to balance the exploration and exploitation. Compared with the model-based approach which builds the model only once and explores the global optimum offline, BO refines the surrogate model incrementally and searches the optimum online. Other successful GP-based approaches have also been proposed \cite{liu2014gaspad,liu2011global}.

In order to further reduce the overall cost of the optimization process, great efforts have been made to exploit the correlations between different fidelity levels of information (\textit{a.k.a.} fidelities) \cite{kennedy2000predicting}. The corresponding multi-fidelity Bayesian optimization algorithms have also been proposed \cite{kandasamy2016gaussian, kandasamy2017multi, poloczek2017multi}. However, only linear correlations are considered in these models and Bayesian optimization algorithms. Recent work \cite{perdikaris2017nonlinear} generalized the linear correlations to general nonlinear one, and proposed nonlinear information fusion algorithms for multi-fidelity modeling. The nonlinear map between the low-fidelity and high-fidelity models provides more flexility to model the complicated correlations between the low- and high-fidelity black-box functions.

Inspired by \cite{perdikaris2017nonlinear}, we propose an efficient multi-fidelity Bayesian optimization approach for analog circuit synthesis. We follow the ideas of \cite{perdikaris2017nonlinear} to build the multi-fidelity model, but propose the weighted Expected Improvement acquisition function and multi-fidelity optimization strategies for Bayesian optimization. GP models are employed to model the low- and high-fidelity black-box functions separately. The nonlinear map between the low-fidelity model and high-fidelity model is also modelled as a Gaussian process. A fusing GP model which combines the low- and high-fidelity models can thus be built. The acquisition function based on the fusing GP model is used to balance the exploitation and exploration. By maximizing the acquisition function and carefully selecting the fidelity for evaluation, our optimization algorithm achieves better optimization results within the limited simulation time. The ability to combine several levels of information to model the slowest one is extremely useful in analog circuit optimization, since we can always carry out the circuit simulation at different precision levels. This would help us to find the global optimums of analog circuits that are impossible to optimize if we only rely on the computationally intensive high-fidelity simulation results. The efficiency of our proposed method has been demonstrated with two real-world analog circuits. And the experimental results show that our proposed methodology can reduce up to 65.5\% of the simulation time during the optimization process.

The remainder of this paper is organized as follow. In section (\S\ref{sec:background}), we present the problem formulation and briefly review the GP-based Bayesian optimization approach. In section (\S\ref{sec:proposal}), the multi-fidelity Bayesian optimization algorithm is proposed. The implementation details are presented in section (\S\ref{sec:details}). And the experimental results is presented in section (\S\ref{sec:experiment}). We conclude the paper in section (\S\ref{sec:conclusion}).

\section{Background} \label{sec:background}


In this section, we first present the problem formulation of analog circuit optimization (\S\ref{sec:problem_formulation}). Then, we give a brief overview of Bayesian optimization (\S\ref{sec:BO}), and the Gaussian process regression (\S\ref{sec:GP}) and the acquisition functions (\S\ref{sec:acquisition_function}).

\subsection{Problem Formulation} \label{sec:problem_formulation}

A typical analog circuit optimization problem can be formulated as a constrained optimization problem, which can be expressed as follows:
\begin{equation}
	\label{eq:Formulation}
    \begin{aligned}
	& \text{minimize} &  & f(\bm{x}) \\
	& \text{s.t.}     &  & c_i(\bm{x}) < 0 \\
	&                 &  & \forall i \in {1 \dots N_c},
    \end{aligned}
\end{equation}
where $\bm{x}$ is a $d$-dimensional vector denoting the $d$ design variables for the analog circuit optimization problem. The $f(\bm{x})$ and $c_i(\bm{x})$ in equation (\ref{eq:Formulation}) represent the objective function and the i-th constraint function respectively. The goal is to minimize $f(x)$ while satisfying the constraints.


\subsection{Overview of Bayesian Optimization} \label{sec:BO}

Bayesian optimization (BO) is a sequential model-based framework that helps to find the global optimum of noisy and expensive black-box functions. Generally, it contains two key ingredients: the surrogate model and the acquisition function \cite{shahriari2016taking}. The probabilistic surrogate model captures our beliefs about the unknown objective function, and provides posterior distribution with both predictions and uncertainties. The acquisition function utilizes the posterior distribution to select the query points sequentially, and it is carefully designed to trade off between exploitation and exploration. The exploitation means the algorithm tends to find the optimum point of the current search space with high confidence. The exploration means the algorithm tries to search the promising unknown area with high predictive uncertainty.

Starting with a randomly sampled training set, Bayesian optimization is able to construct a probabilistic surrogate model to provide the posterior distribution. At each iteration, the training set is incremented by a new point with maximum acquisition function value to refine the surrogate model. After a limited number of iterations, the global optimum is possibly to be found.

\subsection{Gaussian Process Regression} \label{sec:GP}

For Bayesian optimization process, one commonly used surrogate model is the Gaussian process regression (GPR) model \cite{shahriari2016taking}. The GPR model captures our prior beliefs about the objective function and provides the posterior predictive means and well-calibrated uncertainty estimations \cite{rasmussen2004gaussian}.

We consider to approximate a scalar-valued black-box function $f(\bm{x})$ with observation noise $\epsilon \sim N(0, \sigma^2_n)$, where $N(\cdot,\cdot)$ denotes a Gaussian distribution. The training set can be expressed as $D_T = \{X, \bm{y}\}$, where $X = \{\bm{x}_1, \bm{x}_2, ~\dots~, \bm{x}_N\}$ denote the inputs, $\bm{y} = \{y_1, y_2, ~\dots~, y_N\}$ denotes the observations of the black-box function at $\{\bm{x}_1, \bm{x}_2, ~\dots~, \bm{x}_N\}$. We can fully characterize our prior beliefs about the latent function with a Gaussian process whose mean function is $m(\bm{x})$ and kernel function is $k(\cdot,\cdot)$. $k(\cdot,\cdot)$ should be carefully designed to make sure the covariance matrix a symmetric positive definite (SPD) matrix. In this paper, we fix $m(\bm{x})=0$ and select the kernel function as the squared exponential (SE) function, which can be expressed as below
\begin{equation}
	\label{eq:SE}
	k_{SE}(\bm{x}_i,\bm{x}_j) = \sigma^2_f exp(-\frac{1}{2} (\bm{x}_i - \bm{x}_j)^T \Lambda^{-1} (\bm{x}_i - \bm{x}_j)).
\end{equation}
In equation (\ref{eq:SE}), $\Lambda = \text{diag}(l_1,l_2,~\dots~,l_d)$ is a diagonal matrix and $l_i$ denotes the i-th length scale of the i-th dimension. The hyperparameters can be assembled into a single vector $\bm{\theta}=(\sigma_n, \sigma_f, l_1, l_2, ~\dots~, l_d)$ to help optimize the GPR model. By maximize the likelihood, the GPR model can be optimized to approximate the behavior of the expensive latent function $f(\bm{x})$. In other words, we can train the model by minimizing the negative log marginal likelihood (NLML):
\begin{equation}
	\label{eq:NLML}
	NLML = \frac{1}{2} (\bm{y}^T K^{-1}_{\bm{\theta}}\bm{y} + \log \lvert K_{\bm{\theta}} \rvert + N\log(2\pi)),
\end{equation}
where $K_{\bm{\theta}} = K(X,X) + \sigma^2_n I$ is a $N \times N$ covariance matrix.

Given a new input vector $\bm{x}_\ast$, the trained GPR model is capable of providing the corresponding posterior prediction $\mu(\bm{x}_\ast)$ and uncertainty measurement $\sigma^2(\bm{x}_\ast)$. The formulations are as follows:
\begin{equation}
	\label{eq:GP_prediction}
	\begin{cases}
		& \mu(\bm{x}_{\ast}) = k(\bm{x}_{\ast},X)(K+\sigma^2_nI)^{-1}\bm{y} \\
		& \sigma^2(\bm{x}_{\ast}) = \sigma^2_n + k(\bm{x}_{\ast},\bm{x}_{\ast}) - k(\bm{x}_{\ast},X)(K+\sigma^2_nI)^{-1}k(X,\bm{x}_{\ast}),
	\end{cases}
\end{equation}
where $k(\bm{x}_\ast, X) = (k(\bm{x}_\ast, \bm{x}_1), k(\bm{x}_\ast, \bm{x}_2), ~\dots~, k(\bm{x}_\ast, \bm{x}_N))$ and $k(X, \bm{x}_\ast) = k(\bm{x}_\ast, X)^T$.

\subsection{Acquisition Function} \label{sec:acquisition_function}

For simplicity, let us first ignore the constraints and consider the minimization of the objective function. We use weighted expected improvement as acquisition function. Given the current best objective value $\tau$, the improvement can be calculated with $I(y) = max(0, \tau - y)$, and the posterior distribution of $y$ can be provided by the surrogate model $y \sim N(\mu(\bm{x}), \sigma^2(\bm{x}))$. Therefore, the expected improvement (EI) \cite{jones1998efficient} over $\tau$ can be expressed as follows:
\begin{equation}
	\label{eq:EI}
	EI(\bm{x}) = \sigma(\bm{x}) (\lambda \Phi(\lambda) + \phi(\lambda)),
\end{equation}
where $\lambda = (\tau - \mu(\bm{x}))/\sigma(\bm{x})$. In equation (\ref{eq:EI}), $\Phi(\cdot)$ and $\phi(\cdot)$ represent the CDF and PDF of the standard normal distribution respectively. The EI function favors the unknown regions with large uncertainty estimation and the high confidence regions with minimum predictive means.

For constrained optimization problem, there are certain regions that are invalid. The weighted expected improvement (wEI) \cite{schonlau1998global, gardner2014bayesian} algorithm has been proposed to address this problem. By multiplying the EI function with the probability of feasibility (PF), the wEI function favors the feasible regions and reduces the points that are more likely to violate constraints. The corresponding formulation is as below:
\begin{equation}
	\label{eq:wEI}
	wEI(\bm{x}) = EI(\bm{x}) \prod^{N_c}_{i=1} PF_i(\bm{x}),
\end{equation}
where $PF_i(\bm{x})=\Phi(-\mu_i(\bm{x})/\sigma_i(\bm{x}))$ gives a potential measurement of the i-th constraint being satisfied. And $\mu_i(\bm{x})$ and $\sigma_i(\bm{x})$ represent posterior prediction and uncertainty estimation of the i-th constraint function respectively. There are other well-designed acquisition functions, such as upper confidence bounds (UCB) \cite{auer2003using}, predictive entropy search \cite{hernandez2014predictive}, Thompson sampling \cite{chapelle2011empirical}, and knowledge gradient \cite{scott2011correlated}.

\section{Multi-fidelity Bayesian Optimization} \label{sec:proposal}






In this section, we will present our proposed multi-fidelity Bayesian optimization approach for analog circuit synthesis. We will first present the multi-fidelity model and the corresponding training techniques in detail (\S\ref{sec:multi-fidelity_model}). Then, we give the posterior prediction and uncertainty estimation of multi-fidelity model in (\S\ref{sec:multi-fidelity_prediction}). Next, the Bayesian optimization based on the multi-fidelity model is presented in (\S\ref{sec:summary}). In (\S\ref{sec:Fidelity_Selection}), we give a carefully designed fidelity selection criterion.

For simplicity, we only consider modelling technique that involves two levels of fidelities in this paper. We refer the potentially inaccurate but cheap to evaluate model as the low-fidelity model, and the accurate but computation intensive model as the high-fidelity model. We use $f_l(\cdot)$ and $f_h(\cdot)$ denote the GPR model that are trained with the low- and high-fidelity dataset respectively. Given an input vector $\bm{x}$, we use $\mu_l(\bm{x})$, $\sigma^2_l(\bm{x})$, $\mu_h(\bm{x})$ and $\sigma^2_h(\bm{x})$ to represent the corresponding posterior prediction and uncertainty estimation provided by the low- and high-fidelity models respectively. The current best results of coarse and fine data are denoted as $\tau_l$ and $\tau_h$ respectively.

\subsection{Multi-fidelity Model} \label{sec:multi-fidelity_model}


In \cite{kennedy2000predicting}, a multi-fidelity modelling approach which exploits the linear correlations between low-fidelity and high-fidelity models has been proposed as follows:
\begin{equation}
    \label{eq:00multi-fidelity}
    f_h(\bm{x}) = \rho * f_l(\bm{x}) + \delta(\bm{x}),
\end{equation}
where $\rho$ denotes the regression parameter and $\delta(\bm{x})$ represents the independent noise term which follows a Gaussian distribution. This method exhibits great improvement compared with the traditional single-fidelity method. However, it is not suitable for situations when the coarse and fine models exhibit more complex correlations.

Based on the fact that non-functional interplay exists between different fidelity level of information,  a nonlinear information fusion algorithm was proposed in \cite{perdikaris2017nonlinear} to further exploit the corresponding nonlinear cross-correlations between the low- and high-fidelity models.

The first level of the multi-fidelity model is based on the traditional GPR model presented in (\S\ref{sec:GP}). The mean function of the low-fidelity model is fixed as $m(\bm{x}) = 0$ and the corresponding kernel function is the SE function in equation (\ref{eq:SE}). By minimizing the equation (\ref{eq:NLML}), the low-fidelity model trained with the coarse data $D_l=\{X_l, \bm{y}_l\}$ can be optimized to approximate the behavior of the low-fidelity latent function.

In order to enhance the low-fidelity model, the nonlinear correlations between the low- and high-fidelity data should be fully explored. The correlations between low- and high-fidelity models can be generalized as
\begin{equation}
    f_h(\bm{x}) = z(f_l(\bm{x})) + \delta(\bm{x}),
\end{equation}
where $z(\cdot)$ represents the unknown space-dependent correlations between the low- and high-fidelity model, and $\delta(\cdot)$ is a Gaussian process model with respect to the high-fidelity data, which is independent of the low-fidelity model. The nonlinear map $z(\cdot)$ can also be modelled by a Gaussian process model. Since the low-fidelity model $f_l(\bm{x})$ is also a Gaussian process model, $z(f_l(\bm{x}))$ is a deep Gaussian process model, and the posterior distribution of $f_h(\bm{x})$ would thus not be Gaussian. In~\cite{perdikaris2017nonlinear}, the posterior prediction $\mu_l(\bm{x})$ provided by the low-fidelity model $f_l(\bm{x})$ is mapped to the high-fidelity output. With the assumption that the Gaussian processes $\delta(\cdot)$ and $z(\cdot)$ are independent, $f_h(\bm{x})$ is approximated by a Gaussian process with mean function $m_h(\bm{x})=0$ and the kernel function defined as follows \cite{perdikaris2017nonlinear}:
\begin{equation}
    \label{eq:high-fidelity_kernel}
    \begin{aligned}
        k_h(\bm{x}_1,\bm{x}_2; \theta_h) \quad & = \quad k_{h,1}(f_l(\bm{x}_1), f_l(\bm{x}_2); \theta_{h,1}) \\
        \quad & * \quad k_{h,2}(\bm{x}_1, \bm{x}_2; \theta_{h,2}) + k_{h,3}(\bm{x}_1, \bm{x}_2; \theta_{h,3}),
    \end{aligned}
\end{equation}
where $k_{h,1}(\cdot,\cdot)$, $k_{h,2}(\cdot,\cdot)$ and $k_{h,3}(\cdot,\cdot)$ are valid covariance functions, and the squared exponential function in (\ref{eq:SE}) is used in our paper. $\theta_{h,1}, \theta_{h,2}, \theta_{h,3}$ are the corresponding hyperparameters for the kernel functions $k_{h,1}(\cdot,\cdot)$, $k_{h,2}(\cdot,\cdot)$ and $k_{h,3}(\cdot,\cdot)$. We use $\theta_h$ to denote the combined hyperparameters of $\theta_{h,1}, \theta_{h,2}, \theta_{h,3}$. With the $m_h(\bm{x})=0$ and the kernel function $k_h(\bm{x}_1,\bm{x}_2; \theta_h)$ defined in (\ref{eq:high-fidelity_kernel}), the hyperparameters $\theta_h$ can be learnt from the high-fidelity dataset via the minimizing the NLML as shown in (\ref{eq:NLML}). The corresponding posterior distribution is denoted as $f_h(\bm{x}) \sim N(\mu_h(\bm{x}), \sigma^2_h(\bm{x}))$.

\subsection{Prediction and Uncertainty Estimation} \label{sec:multi-fidelity_prediction}

The basic idea of the multi-fidelity model is to use the value of low-fidelity function as additional input parameters, however, if the low fidelity and high fidelity data do not share same input, then $f_l(\bm{x})$ could be unknown, in which case, we model the low fidelity function as Gaussian process and then integrate $f_l(\bm{x})$ out \cite{perdikaris2017nonlinear}. 
\begin{equation}
    \label{eq:multi-fidelity_prediction}
    p(y_{\ast,h}|D,\bm{x}_{\ast}) = \int p(y_{\ast,h}|D_h,y_{\ast,l},\bm{x}_{\ast})p(y_{\ast,l}|D_l, \bm{x}_{\ast}) d\bm{x}_{\ast},
\end{equation}

The posterior distribution of low-fidelity model is Gaussian. The corresponding high-fidelity counterpart is not Gaussian because it depends on the posterior prediction of the low-fidelity model. Given an input vector $\bm{x}_\ast$, the posterior uncertainty estimation of the low-fidelity model will propagate to the higher-fidelity model. With the posterior uncertainty measurement of the low-fidelity model, the overall posterior distribution of the multi-fidelity model can be expressed as follows \cite{perdikaris2017nonlinear}:
where $D_l = \{X_l, \bm{y}_l\}$ and $D_h = \{X_h, \bm{y}_h\}$ correspond to the low- and high-fidelity training data set respectively. $D=\{D_l,D_h\}$ is the combination of the the low- and high-fidelity training data set. $y_{\ast, l}$ and  $y_{\ast, h}$ denote the posterior distribution provided by the low- and high-fidelity model.

\begin{figure}
    \centering
    \includegraphics[width=0.45\textwidth]{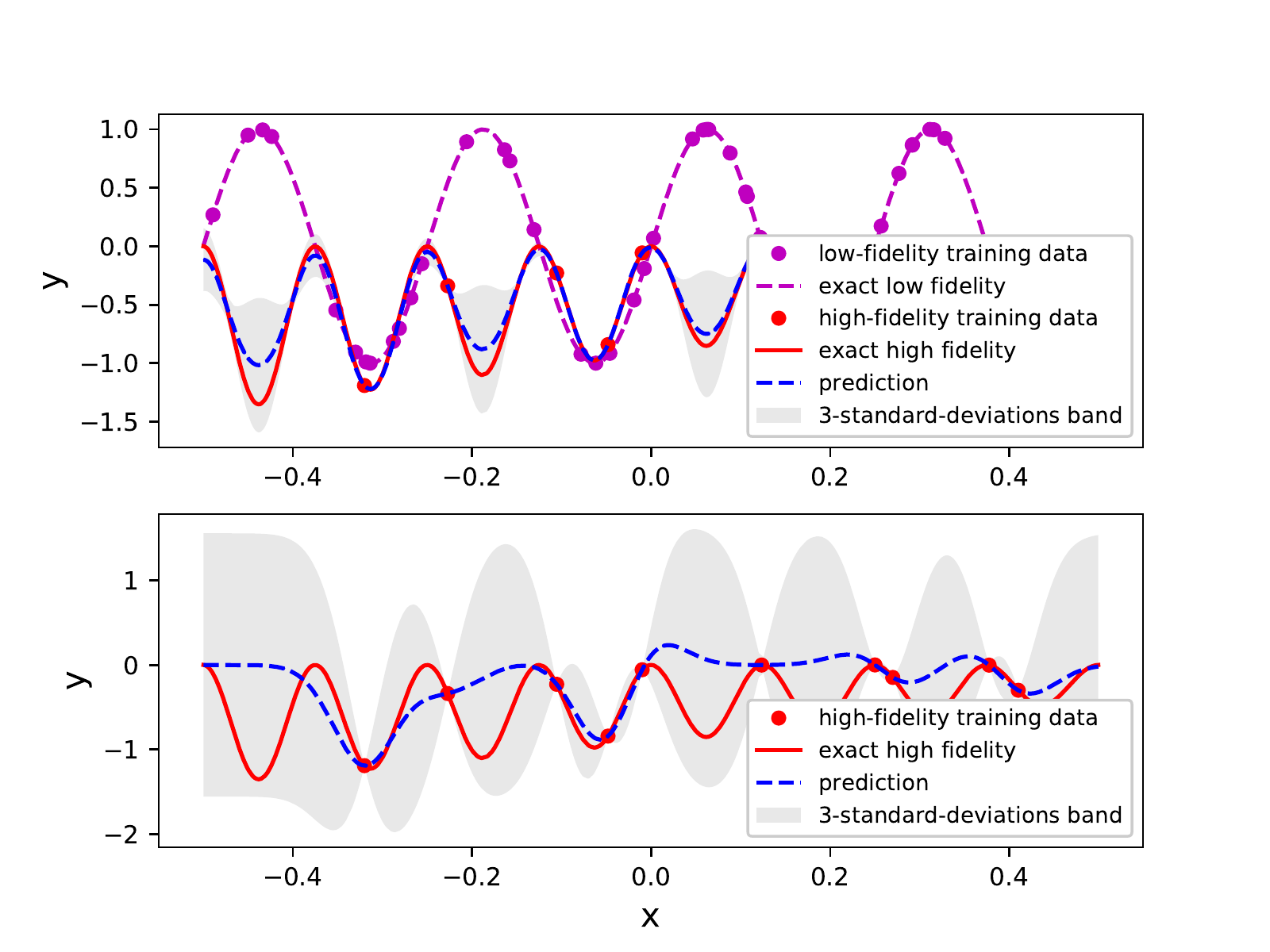}
    \vspace{-0.3cm}
    \caption{The posterior distrbution provided by the multi-fidelity and the traditional single-fidelity GPR model}
    \label{fig:multi-single}
    \vspace{-0.4cm}
\end{figure}

However, computing the posterior distribution with equation (\ref{eq:multi-fidelity_prediction}) faces great challenge. The fundamental problem is that analytically generating the exact posterior predictions and uncertainty estimations is intractable. Therefore, the prediction and uncertainty estimation of the high-fidelity model are obtained through Monte-Carlo integration of (\ref{eq:multi-fidelity_prediction}). We randomly sample $N$ points $X_\ast = (\bm{x}_{\ast,1}, \bm{x}_{\ast,2}, ~\dots~, \bm{x}_{\ast,N})$ that follow the posterior distribution $X_\ast \sim N(\mu_l(\bm{x}), \sigma^2_l(\bm{x}))$, and calculate the prediction and posterior uncertainty with Monte-Carlo process.
It will be used in the subsequent Bayesian optimization process. Figure \ref{fig:multi-single} shows that the posterior prediction fits the latent function better and the uncertainty estimation is much lower compared with the single-fidelity GPR model. The corresponding latent functions come from the pedagogical example of \cite{perdikaris2017nonlinear}.


\subsection{Bayesian Optimization Based on the Multi-fidelity Model} \label{sec:summary}


The proposed multi-fidelity Bayesian optimization approach effectively enhances the low-fidelity model by learning the nonlinear correlations between coarse and fine data automatically. With carefully designed multiple starting point \cite{peng2016efficient,yang2018smart} strategy, the exploitation is greatly improved during the optimization procedure. Fueled with explicitly designed optimization algorithms and the fidelity selection criterion, our proposed method can efficiently reduce the overall optimization cost.

At each iteration, we first construct the low-fidelity GP model and optimize the low-fidelity wEI acquisition function to find an optimal point $\bm{x}^*_l$, then, the high-fidelity GP model is built, and the high fidelity wEI function is optimized based on $\bm{x}^*_l$. The summary of the proposed multi-fidelity Bayesian optimization is given in Algorithm \ref{algo:summary}.

\begin{algorithm}
    \caption{Multi-fidelity Bayesian optimization approach}
    \label{algo:summary}
    \begin{algorithmic}[1]
        \STATE Initialize a training set $D_T=\{X_l,\bm{y}_l,X_h,\bm{y}_h\}$
        \FOR{t = 1 to N}
        \STATE Build the multi-fidelity model
        \STATE Randomly scatter a set of data 
        \STATE Maximize the acquisition function of the low-fidelity model, and get $\bm{x}^*_l$
        \STATE Find $\bm{x}_t$ that maximizes the acquisition function based on $\bm{x}^*_h$
        \STATE Select the fidelity level for evaluation (\S\ref{sec:Fidelity_Selection})
        \STATE Update the training set with the newly sampled data $(\bm{x}_t, y_t)$
        \ENDFOR
    \end{algorithmic}
\end{algorithm}

\subsection{Fidelity Selection Criterion} \label{sec:Fidelity_Selection}
During the optimization process, the surrogate models are incrementally refined with new observed data. By maximizing the acquisition function of the multi-fidelity model, the selected query point can trade off between the exploration and exploitation and reduce the overall optimization cost. Apart from choosing the next query point, we also need select the evaluation fidelity for the multi-fidelity Bayesian optimization process. Based on the fact that the computational cost of sampling the high-fidelity model is much higher than that of the low-fidelity counterpart, we should only sample the high-fidelity data if it is necessary. If the uncertainty of the query point based on the low-fidelity model is low, which means that this data point cannot further improve the accuracy of the low-fidelity model, we will evaluate this query point in high fidelity level to improve the accuracy of multi-fidelity model. Therefore, we select $\bm{x}$ to be sampled in the high-fidelity level if the following criterion is satisfied:
\begin{equation}
    \sigma^2_l(\bm{x}) < \gamma,
\end{equation}
where $\sigma^2_l(\bm{x})$ is the uncertainty estimation with the low-fidelity model, and $\gamma$ is empirically set to be 0.01 in this paper. For constrained optimization problem, the corresponding criterion is expressed as below:
\begin{equation}
    \text{max}(\sigma^2_{l,0}(\bm{x}), \sigma^2_{l,1}(\bm{x}), ~\dots~, \sigma^2_{l,N_c}(\bm{x})) < (1+N_c) * \gamma,
\end{equation}
where $\sigma^2_{l,0}(\bm{x})$ and $\sigma^2_{l,i}(\bm{x})$ are the posterior uncertainty estimation of the objective function and the i-th constraint function provided by the corresponding low-fidelity models. With the proposed fidelity selection criterion, our proposed multi-fidelity Bayesian optimization process would only sample high-fidelity data if it is necessary.

\section{Implementation Details} \label{sec:details}
We will present several implementation details in this section.
\subsection{Multiple Starting Point strategy}

We use multiple starting point \cite{peng2016efficient,yang2018smart} (MSP) strategy to optimize the acquisition function. The region-hit property of the MSP strategy makes it more suitable for global optimization. The stochastic characteristic of MSP algorithm prevents it from getting stuck in the local optimum. By randomly sampling a set of starting point in the solution space, we can expect to cover almost all valleys of the objective function and thus get all local optimum to achieve the global optimum.


\begin{figure}
    \centering
    \includegraphics[width=0.45\textwidth]{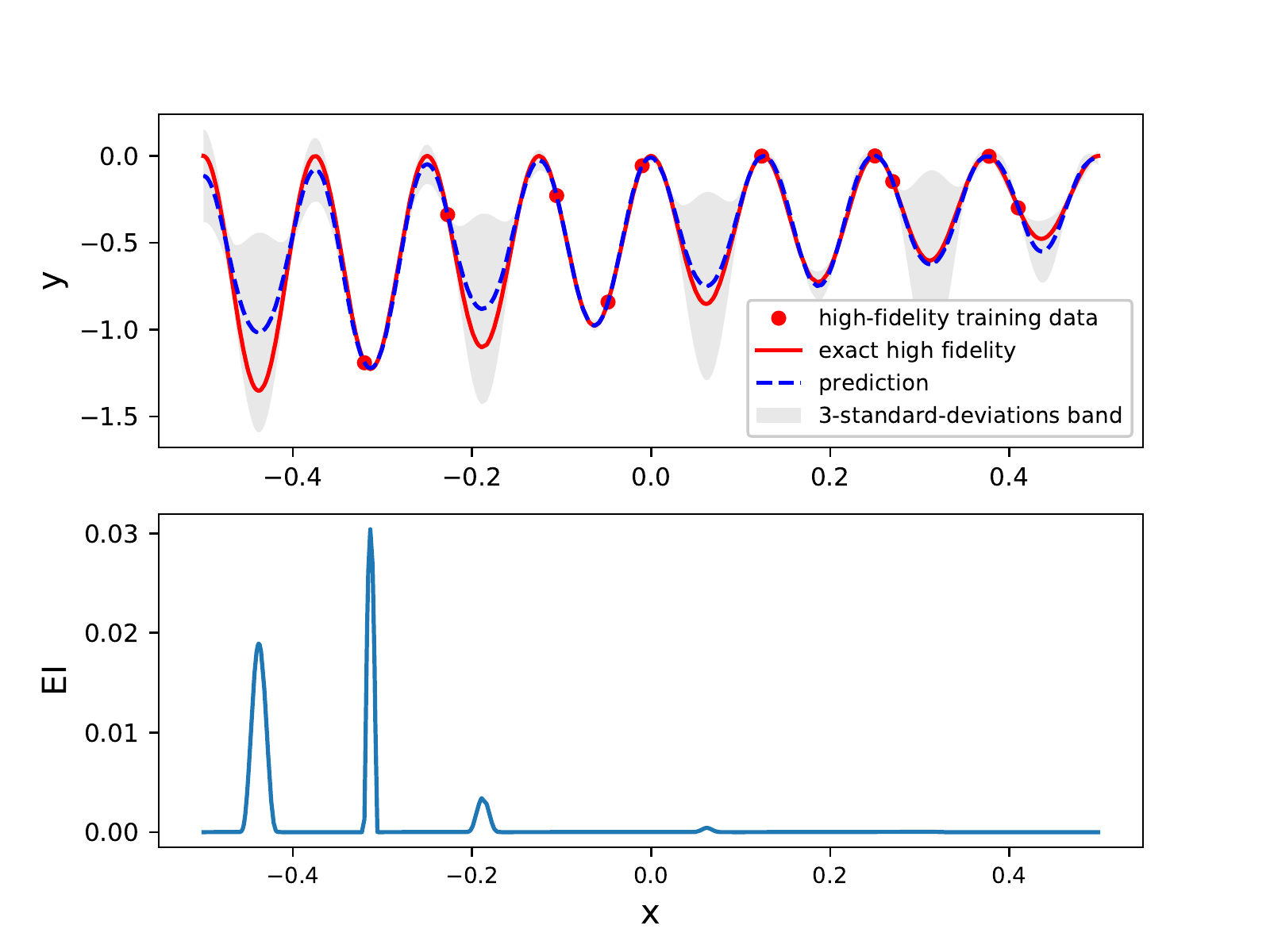}
    \vspace{-0.3cm}
    \caption{The posterior distribution of the multi-fidelity model and the EI function}
    \label{fig:EI}
    \vspace{-0.4cm}
\end{figure}

For the optimization of the acquisition function of the multi-fidelity model, we propose a more sophisticated strategy based on the current search space of the low- and high-fidelity models. During the optimization of the acquisition function of the multi-fidelity model, instead of scattering the starting point randomly, we sample a small fraction of the starting points around $\tau_h$. Considering the fact that the optimum points of the low- and high-fidelity latent function are within a small range in most cases, we also sample a fraction of points around $\tau_l$. By scattering a fraction of points around the best results of the current search space, we can better explore the region that has low posterior predictive mean with high confidence. The reason is that the uncertainty estimation of the current best result is extremely small. And even if there is only a small deviation from the current best result, the gradient of the expected improvement function will be zero as shown in Figure \ref{fig:EI}. The latent functions in Figure \ref{fig:EI} come from the pedagogical example of \cite{perdikaris2017nonlinear}. Therefore, it is not possible to fully explore the current best region with randomly scattered starting points. For constrained optimization problem, where the input dimension is over 20, the problem will be more serious. The ratio at which we sample around $\tau_l$ and $\tau_h$ in this paper is 10 percent and 40 percent respectively.

\subsection{Finding the First Feasible Solution}

For the constrained optimization problem, there is a possibility that the initial dataset has no feasible points. Considering the narrow bell shape of both EI and PF function, we minimize the sum of the posterior means which violate the constraints to force the starting point into a feasible region. The corresponding formulation is as follow:
\begin{equation}
    \label{eq:sqp_stage1}
    \text{minimize} \quad \sum^{n_c}_{i=1} \text{max}(0,\mu_{h,i}(\bm{x})).
\end{equation}
This operation helps to accelerate the process of finding the first feasible point. And it further reduces the computational resources spent on searching the infeasible region during the optimizaion process.



\section{Experimental Results} \label{sec:experiment}

In this section, we demonstrate the efficiency of our proposed optimization approach with two real-world analog circuits: power amplifier (\S\ref{sec:PA}) and charge pump (\S\ref{sec:charge_pump}). We quantitatively compare our approach with three state-of-the-art algorithms: WEIBO \cite{lyu2018efficient}, GASPAD \cite{liu2014gaspad} and DE \cite{liu2009analog}. The WEIBO method is a traditional single-fidelity GP-based BO approach with wEI function works as the acquisition function. The GASPAD approach is a GP-based approach with evolutionary algorithm works as the optimization engine and lower confidence bound \cite{dennis1997managing,emmerich2006single} works as the acquisition function. The DE methodology is an optimization approach based on the evolutionary algorithm. All our experiments are conducted on a Linux workstation with two Intel Xeon CPUs and 128G memory.

\subsection{Power Amplifier} \label{sec:PA}

The power amplifier circuit is implemented in a TSMC 65nm process and works at a frequency of 2.4GHz. The array-based power amplifier contains 2048 duplicated cells, and they have 4 transistors each. We aim to maximize the output efficiency (Eff) of this circuit while trying to meet two constrains, which are the output power (Pout) and the total harmonic distortion (thd). The corresponding design specifications are as follows:
\begin{equation}
    \begin{aligned}
    \text{maximize} \quad &  \mathit{Eff} \quad &  \quad &\\
    \text{s.t.} \quad & \mathit{Pout} \quad & > \quad & 23dBm \\
                \quad & \mathit{thd} \quad & < \quad & 13.65dB.
    \end{aligned}
\end{equation}

\begin{figure}
    \centering
    \includegraphics[width=0.40\textwidth]{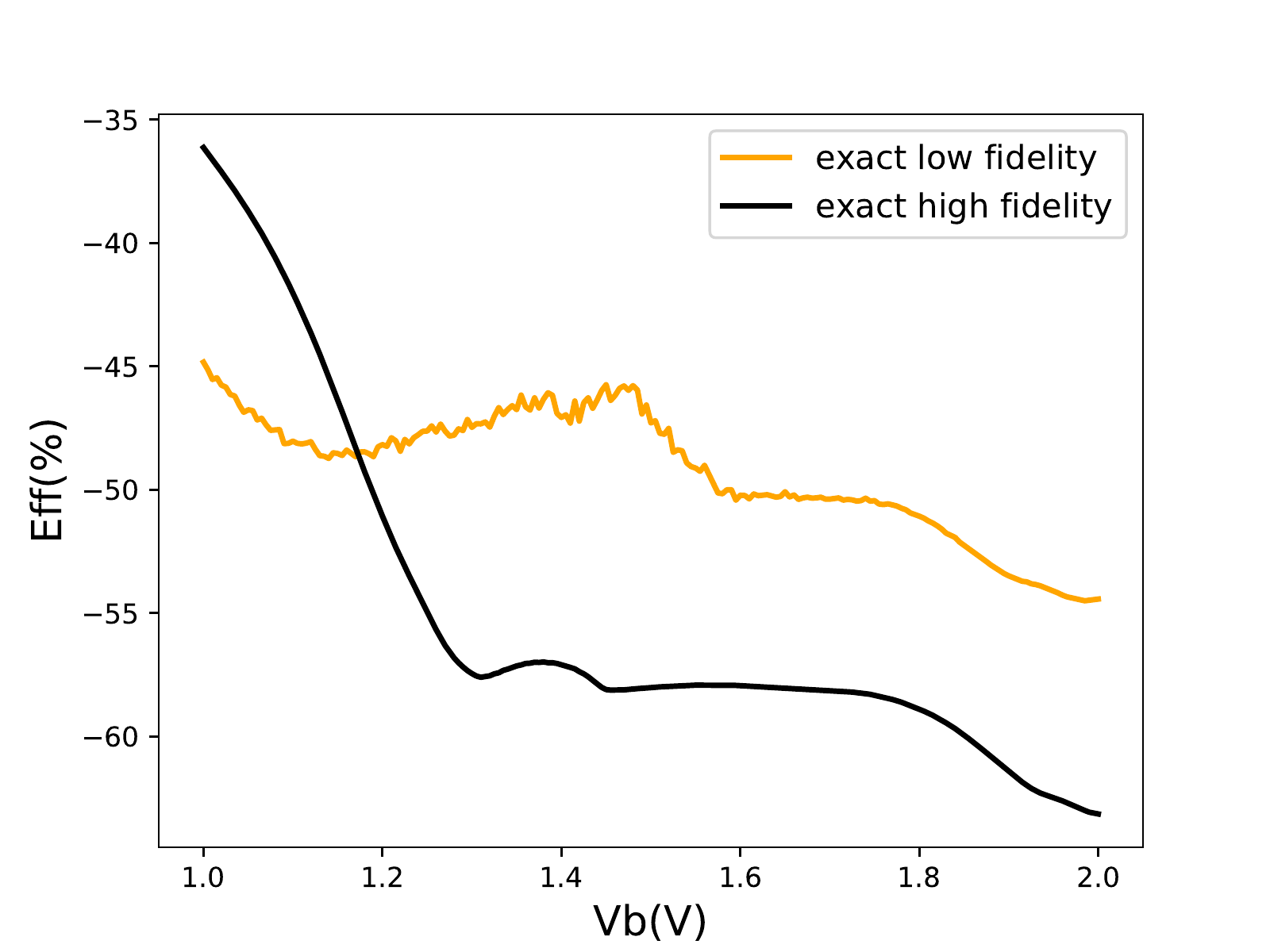}
    \vspace{-0.3cm}
    \caption{Nonlinear correlations between the low- and high-fidelity simulation results}
    \label{fig:PA}
    \vspace{-0.4cm}
\end{figure}

There are 5 design variables in the power amplifier circuit. Figure \ref{fig:PA} fixes four design variables $C_s, C_p, W, V_{dd}$ and presents the low- and high-fidelity simulation results. It shows that nonlinear cross-correlations exist between the low- and high-fidelity dataset. In this experiment, we set the simulation time of each transistor as 10ns and 200ns for low- and high-fidelity model respectively. As the evaluation time of high-fidelity model is much higher than that of the low-fidelity one, we limit the high-fidelity data budget. For our proposed method, the budget is set to be 150 and initialized with 10 low-fidelity data and 5 high-fidelity data. WEIBO is initialized with 40 high-fidelity data points and limited with 150 simulations. For both GASPAD and DE, the simulation budget is set to be 300. We run our proposed method 12 times to average out the random fluctuations. The experimental results are shown in Table \ref{table:PA}. On average, our proposed method requires 252 coarse data and 46 fine data to reach the corresponding results, which is equivalent to the simulation time of 59 high-fidelity data. Although the best optimization result of WEIBO is slightly better than ours by 0.6\%, our proposed method reduces about 28.0\% of the simulation time and exhibits more stable performance. Compared with GASPAD and DE which also achieve feasible design, our proposed method achieves much better results with less simulation time. The experimental results also show that our proposed method is still competitive even if the evaluation cost has already been reduced to a relatively low level.



\begin{table}
    \centering
    \caption{The optimization results of the power amplifier, the results of WEIBO, DE, GASPAD come from \cite{lyu2018efficient}}
    \label{table:PA}
    \vspace{-0.2cm}
    \begin{tabular}{ccccc}
        \hline
        \textbf{Algo} & \textbf{Ours} & \textbf{WEIBO} & \textbf{GASPAD} & \textbf{DE} \\
        \hline
        \text{thd/dB}         & 7.40 & 8.87  & 13.05 & 12.39  \\
        \text{Pout/dBm}       & 23.05 & 23.11 & 24.23 & 24.25  \\
        \hline
        \text{Eff(mean)/\%}   & \textbf{62.64} & 60.29 & 31.63 &  31.54 \\
        \text{Eff(median)/\%} & \textbf{63.16} & 62.23 & 22.58 &  33.35 \\
        \text{Eff(best)/\%}   & 63.65 & \textbf{64.02} & 62.02 &  53.69 \\
        \text{Eff(worst)/\%}  & \textbf{60.84} & 48.86  & 22.07 &  14.07 \\
        \hline
        Avg. \# Sim & \textbf{59} & 82 & 257 & 234 \\
        \# Success & \textbf{12/12} & 12/12 & 12/12 & 12/12 \\
        \hline
    \end{tabular}
    \vspace{-0.3cm}
\end{table}

\begin{table}
    \centering
    \caption{The Optimization results of the charge pump}
    \label{table:charge_pump}
    \vspace{-0.2cm}
    \begin{tabular}{ccccc}
        \hline
        \textbf{Algo} & \textbf{Ours} & \textbf{WEIBO} & \textbf{GASPAD} & \textbf{DE} \\
        \hline
        $\rm max\_diff_{1}$ & 6.26 & 6.44 & 6.58 & 9.37 \\
        $\rm max\_diff_{2}$ & 3.7 & 4.83 & 4.86 & 6.9 \\
        $\rm max\_diff_{3}$ & 0.16 & 0.14 & 0.28 & 0.22 \\
        $\rm max\_diff_{4}$ & 0.37 & 0.16 & 0.46 & 0.39 \\
        \text{deviation} & 0.74 & 0.31 & 0.41 & 0.23 \\
        \hline
        \text{mean} & \textbf{3.99} & 4.23 & 4.22 & 5.88 \\
        \text{median} & \textbf{3.96} & 4.20 & 4.17 & 5.81 \\
        \text{best} & \textbf{3.52} & 3.63 & 3.86 & 5.18 \\
        \text{worst} & \textbf{4.38} & 4.83 & 4.79 & 6.39 \\
        \hline
        Avg. \# Sim & \textbf{158} & 458 & 2177 & 9499 \\
        \# Success & \textbf{10/10} & 10/10 & 10/10 & 10/10 \\
        \hline
    \end{tabular}
    \vspace{-0.3cm}
\end{table}

\subsection{Charge Pump} \label{sec:charge_pump}

\begin{figure}
    \centering
    \includegraphics[width=0.46\textwidth]{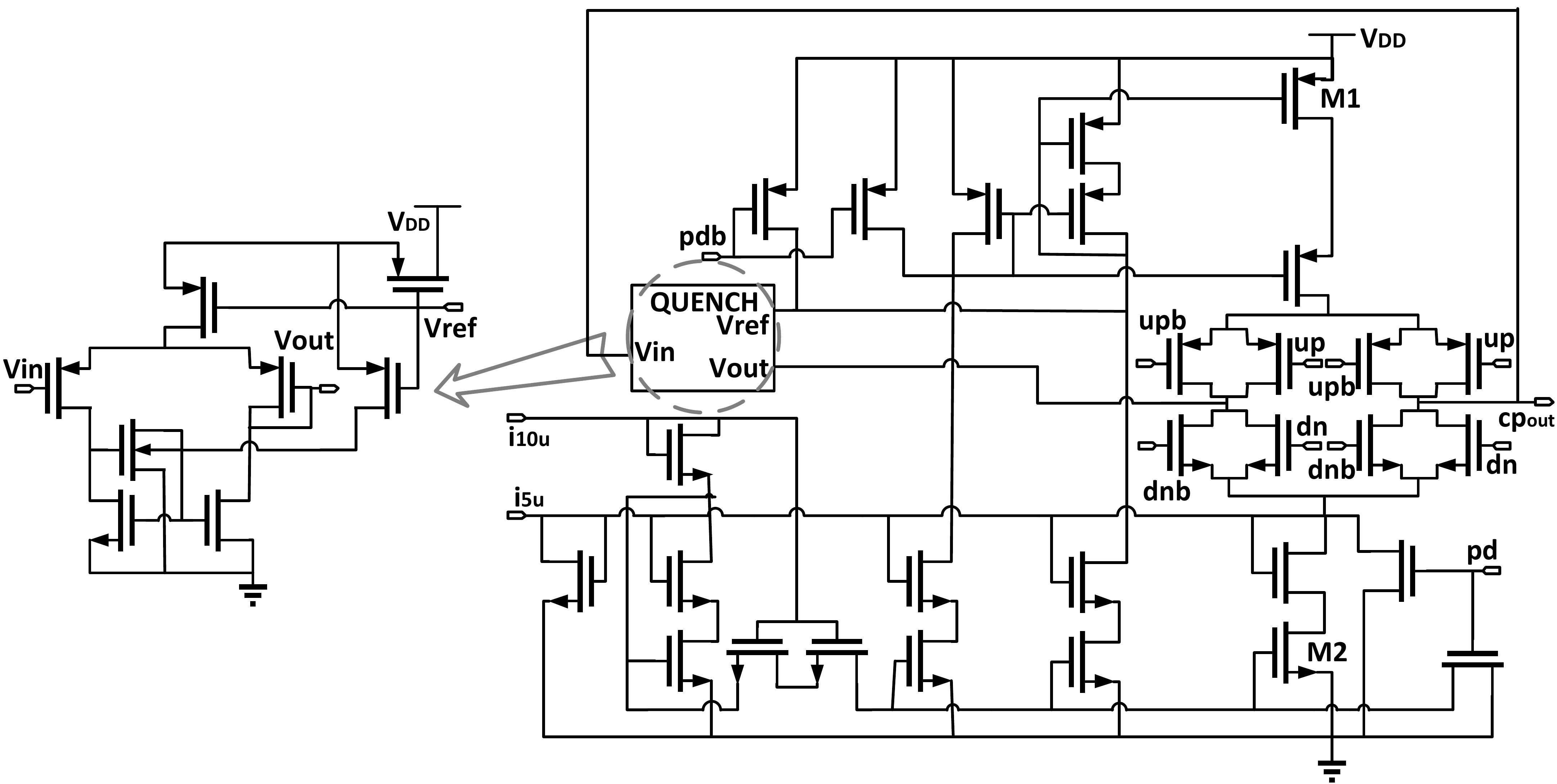}
    \vspace{-0.3cm}
    \caption{Schematic of the charge pump, reproduced from \cite{yang2018smart}}
    \label{fig:charge_pump}
    \vspace{-0.4cm}
\end{figure}

The schematic of the charge pump circuit is shown in Figure \ref{fig:charge_pump}, which is designed with a SMIC 40nm process and has 36 design variables. The aim of charge pump circuit design is to contrain the current of transistor $M_1$ and $M_2$ within a small range around 40$\mu$A for simulations over a total of 27 PVT corners. The low-fidelity model of our proposed method is instead simulated with only a single PVT corner. During the simulation process, we evaluate the maximum, average and minimum current for both transistor $M_1$ and $M_2$. The corresponding design specifications are as follows:
\begin{equation}
    \begin{aligned}
        \text{minimum} \quad & FOM & & \\
        \text{s.t.} \quad & \mathit{max\_diff_1} \quad & < \quad & 20\mu A \\
                    \quad & \mathit{max\_diff_2} \quad & < \quad & 20\mu A \\
                    \quad & \mathit{max\_diff_3} \quad & < \quad & 5\mu A \\
                    \quad & \mathit{max\_diff_4} \quad & < \quad & 5\mu A \\
                    \quad & \mathit{deviation} \quad & < \quad & 5\mu A,
    \end{aligned}
\end{equation}
where
\begin{equation}
    \label{eq:charge_pump}
    \begin{cases}
    \begin{aligned}
        \mathit{max\_diff_1} \quad & = \quad \text{max}_{\forall PVT}(I_{M_1,max} - I_{M_1,avg})\\
        \mathit{max\_diff_2} \quad & =  \quad \text{max}_{\forall PVT}(I_{M_1,avg} - I_{M_1,min})\\
        \mathit{max\_diff_3} \quad & =  \quad \text{max}_{\forall PVT}(I_{M_2,max} - I_{M_2,avg})\\
        \mathit{max\_diff_4} \quad & = \quad \text{max}_{\forall PVT}(I_{M_2,avg} - I_{M_2,min})\\
        \mathit{deviation} \quad & = \quad \text{max}_{\forall PVT}(\lvert I_{M_1,avg}-40\mu A\rvert) \\
        \quad & + \quad \text{max}_{\forall PVT}(\lvert I_{M_2,avg}-40\mu A\rvert) \\
        FOM \quad & = \quad 0.3 \times \sum^4_{i=1} \mathit{max\_diff_i} + 0.5 \times \mathit{deviation}.
    \end{aligned}
    \end{cases}
\end{equation}

In order to average out the random fluctuations, we run each algorithms 10 times. As the evaluation cost of high-fidelity model is much more expensive than the low-fidelity one, we set the high-fidelity simulation budget as 300 and randomly initialize the training set with 30 low-fidelity data and 10 high-fidelity data. Both WEIBO and GASPAD are initialized with 120 data points and the maximum number of simulations are limited as 800 and 2500 respective. For DE, the simulation budget is set to be 10100 with 100 initial data points. The experimental results are presented in Table \ref{table:charge_pump}. 
On average, our proposed method requires 325 coarse data and 146 fine data to achieve the corresponding results, which translates to the simulation time of 158 fine data. Compared with WEIBO, our proposed method reduces approximately 65.5\% of the simulation time while achieve better optimization results. Although both GASPAD and DE also achieve the feasible design, the overall optimization cost is much higher than our approach. The results of the charge pump circuit also show that our method still works well when the input dimension is over 20.





\section{Conclusion} \label{sec:conclusion}

In this paper, we proposed an efficient multi-fidelity Bayesian optimization approach for analog circuit synthesis. By considering the nonlinear cross-correlations between the low- and high-fidelity data, our proposed method is capable of enhancing the low-fidelity model with only a few high-fidelity data injected. Fueled by the explicitly designed optimization algorithm and fidelity selection criterion, our approach demonstrates competitive optimization performance. Compared with the state-of-the-art single-fidelity optimization approach, our proposed method can reduce up to approximately 65.5\% of the simulation time.

\section*{Acknowledgements}
This research is supported partly by the National Major Science
and Technology Special Project of China (2017ZX01028101-
003), partly by National Natural Science Foundation of China
(NSFC) research projects 61822402, 61774045, 61574046,
61674042 and 61574044.

\bibliographystyle{ACM-Reference-Format}
\bibliography{bibliography}

\end{document}